\documentclass[letterpaper]{article}

\usepackage{natbib,alifeconf}  
\usepackage{hyperref}

\title{A model of urban evolution based on innovation diffusion}
\author{Juste Raimbault$^{1,2,3}$\\
\mbox{}\\
$^1$Center for Advanced Spatial Analysis, University College London\\
$^2$UPS CNRS 3611 ISC-PIF\\
$^3$UMR CNRS 8504 G{\'e}ographie-cit{\'e}s\\
juste.raimbault@polytechnique.edu} 

\begin{document}
\maketitle

\begin{abstract}
The dynamics of urban systems can be understood from an evolutionary perspective, in some sense extending biological and cultural evolution. Models for systems of cities implementing elementary evolutionary processes remain however to be investigated. We propose here such a model for urban dynamics at the macroscopic scale, in which the diffusion of innovations between cities captures transformation processes (mutations) and transmission processes (diffusion), using two coupled spatial interaction models. Explorations of the model on synthetic systems of cities show the role of spatial interaction and innovation diffusion ranges on measures of diversity and utility, and the existence of intermediate ranges yielding an optimal utility. Multi-objective optimization shows how the model produces a compromise between utility and diversity. This model paves the way towards more elaborated formalizations of urban evolution.
\end{abstract}

\section{Introduction}


Urban systems are complex as they combine technical artefacts with socio-economic dynamics at multiple temporal and spatial scales. An understanding of processes driving their dynamics is an important aspect for shaping sustainable policies, and a sustainable design and management of cities \citep{lobo2020urban}. Multiple disciplines and viewpoints have proposed such insights from a complexity perspective \citep{pumain2020conclusion}, and Artificial Life approaches have shown promising results to study urban systems, both through conceptual contributions such as interpreting cities through biological metaphors \citep{batty2009centenary}, but also through modeling and simulation for urban growth including cellular automata and evolutionary computation \citep{raimbault2020cities}.


A stream of research tightly linked to ALife relates to concepts of \emph{Urban Evolution}. These build in a sense on Cultural Evolution \citep{mesoudi2001cultural}, which aims at understanding changes in social knowledge as evolutionary processes involving replication, mutation, selection. It bears similarities with biological evolution but is not reducible to it, and for example uses the concept of \emph{meme} as transmission units comparable to genes. Both biological and cultural evolution can be linked into common frameworks and models, implying the coupling of different timescales \citep{bull2000meme}. In terms of urban studies, the concept of urban evolution is less formalized than cultural evolution and can be understood in multiple ways. \cite{votsis2019urban} use the concept of Urban DNA to characterize morphological properties of cities such as population density or the role of the road network. Similarly, \cite{kaya2017urban} describe cities based on their morphological properties as a product of their dynamics. \cite{wu2011urban} link the parameters of a cellular automaton model of urban growth to intrinsic properties of urban regions, which should play a role in their overall evolution. \cite{d2014urban} propose an urban genetic code as the way agents cooperate and compete, cities emerging as dynamical equilibria from these interactions. In Economic Geography, the concept of co-evolution is used mostly for urban agents such as firms and stakeholders \citep{gong2019co}. At the macroscopic scale of urban systems, \cite{pumain2018evolutionary} proposes an evolutionary theory to study systems of cities as complex adaptive systems, in which interactions between cities play a crucial role. Corresponding simulation models were proposed with different thematic focus, for example by \cite{cottineau2015modular} with economic exchanges and by \cite{raimbault2020indirect} with infrastructure networks. In Urban Design, \cite{batty2009digital} introduce evolutionary computation to explore possible urban forms. There still however remains a lack of models which would operationalize the concept of urban evolution in a way close to its biological and cultural counterparts, i.e. capturing explicitly the fundamental processes of transmission and transformation within differentiating subsystems \citep{durham1991coevolution}.


Besides, a central concept bridging ALife and evolution with the study of urban systems is the concept of innovation diffusion. Within artificially evolved systems, understanding how innovation emerges and how it diffuses in the population is essential \citep{bedau2000open}. One aspect of open-ended evolution are mechanisms endogeneously producing novelty. In the case of urban systems, \cite{hagerstrand1968innovation} had already highlighted the role of a hierarchical diffusion of innovation between cities in their trajectories. The aforementioned evolutionary urban theory suggests that innovation cycles and their hierarchical diffusion is a possible explanation for the properties of urban scaling laws \citep{pumain2006evolutionary}. Evolutionary economics also study regional systems of innovation, how it diffuses and thee potential existence of spatial spillovers \citep{uyarra2010evolutionary}. Thus, the diffusion of innovation is a privileged entry into urban evolutionary processes.


This paper proposes to tackle the issue of modeling urban evolution by using innovation diffusion processes to capture elementary evolutionary processes. We investigate thus the question to what extent simple models of urban evolution integrating an urban genome can be used to simulate urban dynamics. Our contribution relies on the following points: (i) we describe a relatively simple model for systems of cities at the macroscopic scales, based on population growth and the diffusion of innovation between cities; (ii) we systematically explore this model on synthetic systems of cities to extract stylized facts on different indicators including global diversity and utility.

The rest of this paper is organized as follows: in the next section we give the model context and describe it formally. We then present numerical experiments, including internal statistical validation, exploration of the parameter space, and optimization using a multi-objective genetic algorithm. We finally discuss implications of our results and possible future extensions.

\section{Urban evolution model}

\subsection{Rationale}

The core idea of the model is to build on a concept of ``Urban DNA'' which would capture evolution processes as in biological evolution and cultural evolution, i.e. a kind of genome that cities would be exchanging and which would undergo mutation processes. A suitable candidate is to build on the concept of \emph{meme} introduced in the field of cultural evolution. However, several particularities must be stressed out when working with urban systems. First, they are multi-dimensional implying very different types of agents, including physical and technical artefacts (e.g. infrastructures) but also social and economical structures. A comprehensive urban genome would include these very different dimensions. Then, they are multi-scalar in time and space, meaning that evolution processes, if they exist, may occur at different paces and through different elementary carriers. Finally, they are embedded in the geographical space, what structures the way transmission and mutation can occur, but also can strongly change properties of underlying processes \citep{raimbault2019space}. We choose to focus on the last point, considering a model at the macroscopic scale where cities are agents and with a one-dimensional genome, but in which spatial interactions are crucial for the dynamics. More particularly, this model combines two spatial interaction models \citep{fotheringham1989spatial}.

Several models have been proposed to simulate the diffusion of innovation at microscopic and mesoscopic scales \citep{kiesling2012agent}. \cite{blommestein1987adoption} describe a model of innovation diffusion and urban dynamics with endogenous demand for innovations, but in which the spatial component only influences prices of innovations. \cite{deffuant2005individual} give an example of an elaborated model for adoption dynamics at the microscopic level. Effective channels for the diffusion of innovations are multiple, and can for example be urban firm linkages \citep{rozenblat2007firm}. \cite{pumain2017simpoplocal} describe a model based on innovation diffusion to explain the emergence of the first cities. \cite{favaro2011gibrat} study an urban growth model including innovation diffusion at the scale of the urban system. Our model builds on this last framework, extending and adapting it to an urban evolution context. It is however significantly different in the way population are updated deterministically and in the innovation process.

In our model, cities are characterized by their size in terms of population, and city size evolve following a spatial interaction model in which city attractivity is included. This attractivity is determined by how innovative cities are. Transformation processes are included as mutations, when random innovation appear in cities. Transmission processes (spatial crossover) are included by diffusing innovations between cities. Finally, to existence of subsystems in which evolution can occur is natural through the spatial aspect of the model, through which different regions in space may behave and evolve differently, possibly resulting in the emergence of co-evolution niches \citep{holland2012signals,raimbault2018co}.

\subsection{Model description}

 
We now formally describe the urban evolution model. A number $N$ of cities are located in the geographical space, and described in time by their size $P_i (t)$ (which generally corresponds to population). The geography is captured with a distance matrix $d_{ij}$ between cities. We consider a one-dimensional technological space in which innovations can be introduced, indexed by their order of apparition $c$. Cities are then also characterized by their innovation profile $\delta_{c,i} (t) \in \left[0;1\right]$ which represents the proportion of population having adopted innovation $c$ in the city $i$ at time $t$. We assume exclusivity in the adoption of innovation, i.e. that we always have $\sum_c \delta_{c,i} (t) = 1$. This innovation profile corresponds to the urban genome. Each innovation has furthermore an utility $u_c >0$ which will influence its diffusion dynamics.

\paragraph{Model dynamics}


Starting from an initial configuration, the simulation model is iterated in time with the following steps at each time tick:

\begin{enumerate}
	\item The crossover between urban genomes relies on spatial processes of innovation diffusion. This means that existing innovations are propagated between cities following a spatial interaction model given by

\begin{equation}
\delta_{c,i,t} = \frac{\sum_j p_{c,j,t-1}^{\frac{1}{u_c}} \cdot \exp{(-\frac{d_{ij}}{d_I})}}{\sum_c \sum_j p_{c,j,t-1}^{\frac{1}{u_c}} \cdot \exp{(-\frac{d_{ij}}{d_I})}}
\end{equation}

where $p_{c,i,t} = \delta_{c,i,t} \cdot \frac{P_{i}(t)}{\sum_k P_k (t)}$ is the share of total population adopting $c$ in the city $i$ at time $t$; $d_I$ is the characteristic distance in the spatial innovation model, for which an increase will correspond to a broader spatial diffusion. As the population share is between 0 and 1, higher utilities will effectively diffuse faster as the exponent $1/u_c$ is used. In other terms, the inverse of the utility is the coefficient of population in the underlying spatial interaction model, and using an exponent is more relevant in that sense.

	\item The sizes of cities evolve according to their performance in terms of innovation, i.e. more innovative cities are more attractive. This stage is also deterministic, following  with $P_i(t) - P_i(t-1) = w_I\cdot \sum_j \frac{V_{ij}}{<V_{ij}>}$ where $w_I$ is a fixed growth rate fixed to $w_I = 0.005$ in experiments, and where the interaction potential is defined by
\begin{equation}
V_{ij}= \frac{P_{i}(t-1) \cdot P_{j}(t-1)}{(\sum_k P_k(t-1))^2} \cdot \exp{\left(-\frac{d_{ij}}{d_G} \cdot \prod_c \delta_{c,i,t}^{\phi_{c,t}}\right)}
\end{equation}
where $\phi_{c,t} = \sum_i \delta_{i,c,t}\cdot P_i(t-1) /\sum_{i,c} \delta_{i,c,t}\cdot P_{i}(t-1)$ is the macroscopic adoption level (globally more adopted innovations will have a higher attractivity); $d_G$ is the characteristic distance for this second spatial interaction model. This model for population dynamics was proposed in such a setting by \citep{raimbault2020indirect} without the additional innovation attractivity term.

 \item Finally, mutations in the urban genome (transformation process) corresponds to the introduction of new innovations. Each city has an innovation activity $\beta$ independent of its size, which corresponds to an intrinsic mutation rate; and will have a probability to innovate function of its size, similar to a Gibrat model \citep{pumain2012theorie}, as $\beta \cdot \left(P_i (t) / \max_k P_k (t)\right)^{\alpha_I}$ where $\alpha_I$ is a hierarchy exponent biasing the innovation towards larger cities. The utility for a new innovation is drawn stochastically, following either a normal or log-normal distribution, such that (i) its average corresponds to the current mean of utilities for existing innovations, and (ii) its standard deviation is a fixed parameter $\sigma_U$, which allows controlling the ``disruptivity'' of innovations. Urban genomes are adapted such that the new innovation as an initial penetration rate $r_0$ in the innovative city (previous innovation shares are rescaled for this city).
\end{enumerate}

The urban evolution is stopped after a fixed number of steps $t_f$.

\paragraph{Synthetic setup}

Although \cite{raimbault:halshs-01880492} proposed to apply the model of \cite{favaro2011gibrat} on real systems of cities, we focus our model study here on synthetic systems of cities, in order to isolate intrinsic effects independently of geographical contingencies. Cities are located randomly in an uniform square space. City sizes follow a Zipf law with exponent $\alpha_0 = 1$, i.e. such that $P_i (0) = \frac{P_{max}}{i^{\alpha_0}}$ \citep{pumain2012theorie}. We do not modify this initial hierarchy in our experiments. Distances between cities $d_{ij}$ are computed as euclidian distances. We take $N=30$ cities and $t_f = 50$ in our experiments, corresponding to a regional or national system of cities, on timescales of the order of a century.

\paragraph{Indicators}


The behavior of the model is quantified through indicators at the macroscopic scale. These should allow extracting stylized facts from the model exploration. We consider the following indicators:
\begin{itemize}
	\item Average diversity, defined as an Herfindhal diversity index over innovation shares across cities, averaged in time as
	\begin{equation}
		D = \frac{1}{t_f + 1} \sum_{t=0}^{t_f} \left(1 - \sum_{i,c} \left(p_{c,i,t}\right)^2 \right)	
	\end{equation}
	Note that other diversity indices could be applied, or a similar be computed on macroscopic adoption shares or within each city, and would yield different results. This one has the advantage to combine the diversity within and across cities.
	\item Average utility, given by the weighted average of innovation utilities, averaged in time, as
	\begin{equation}
		U = \frac{1}{t_f + 1} \sum_{t=0}^{t_f} \sum_{i,c} \delta_{c,i,t} u_c 
	\end{equation}
	\item Innovativity, given by the average number of innovations per city and unit of time $I = \frac{\max c}{N\cdot (t_f + 1)}$.
	\item To quantify the trajectories of city populations, many indicators can be used \citep{raimbault2020unveiling}. We choose simply to quantify the level of hierarchy at final time, estimated by fitting the rank-size law on populations using an Ordinary Least Squares on logarithms. This captures if the urban systems has become more unequal in terms of population balance.
\end{itemize}

\begin{figure*}[t]
	\centering
	\includegraphics[width=0.49\linewidth]{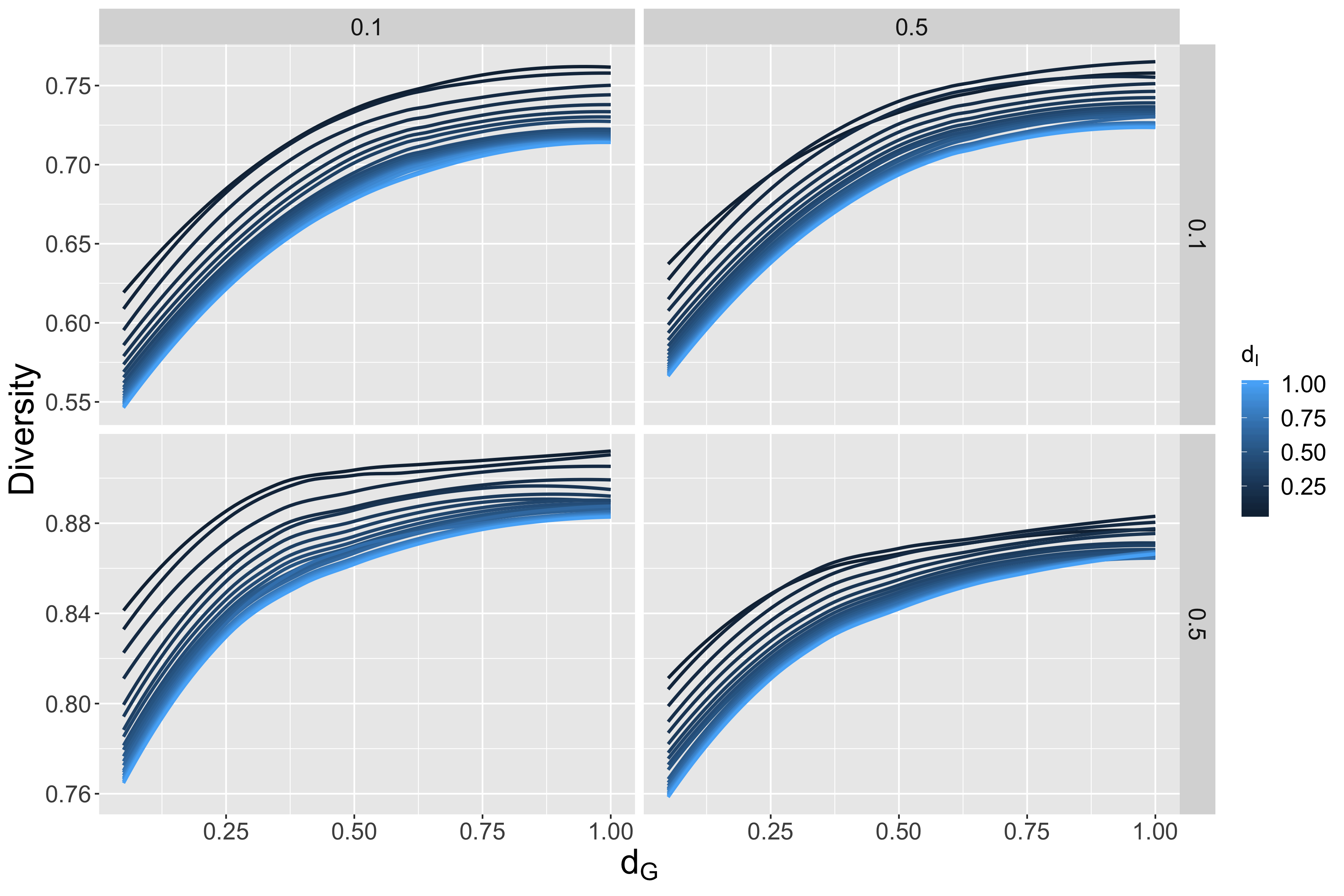}
	\includegraphics[width=0.49\linewidth]{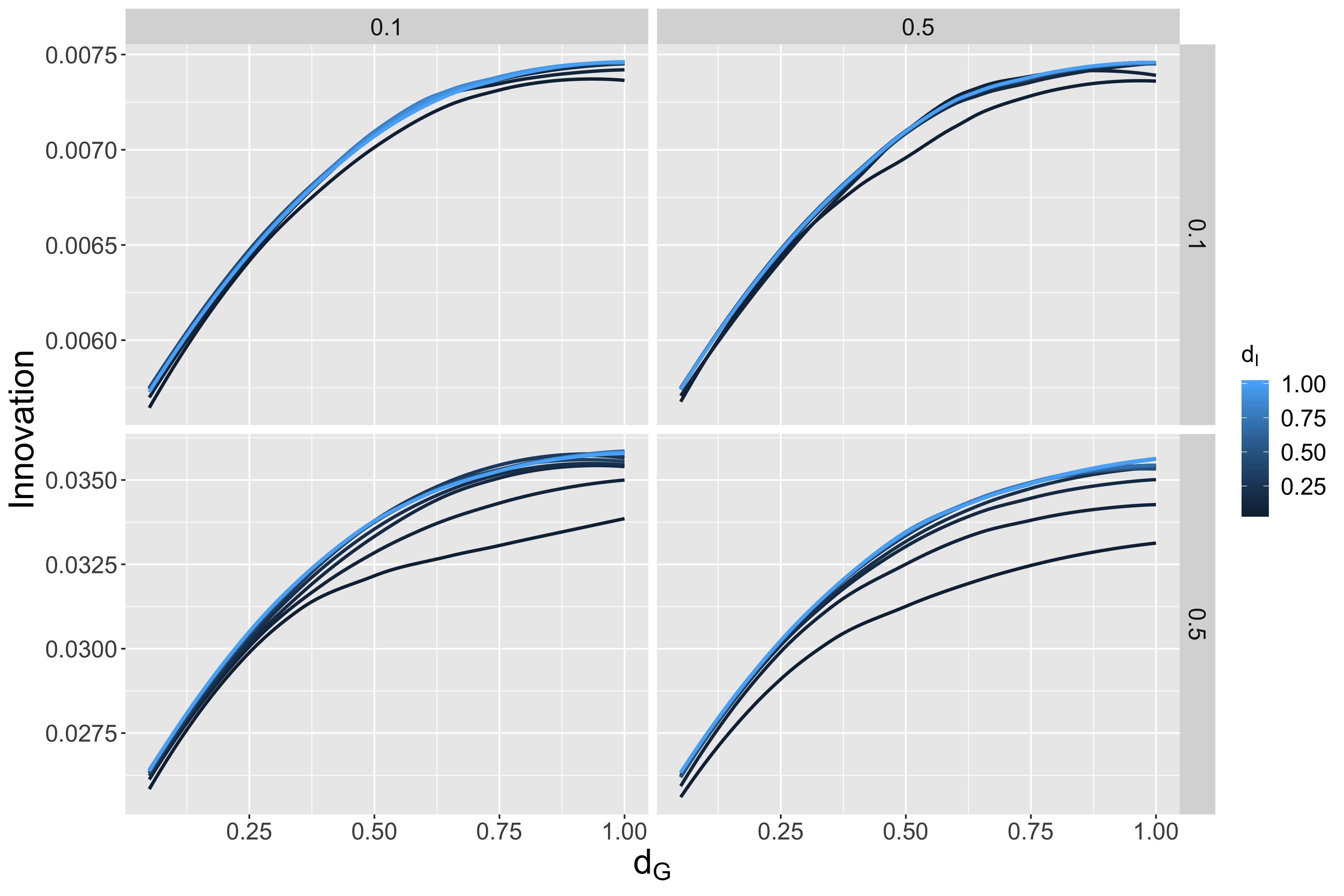}
	\includegraphics[width=0.49\linewidth]{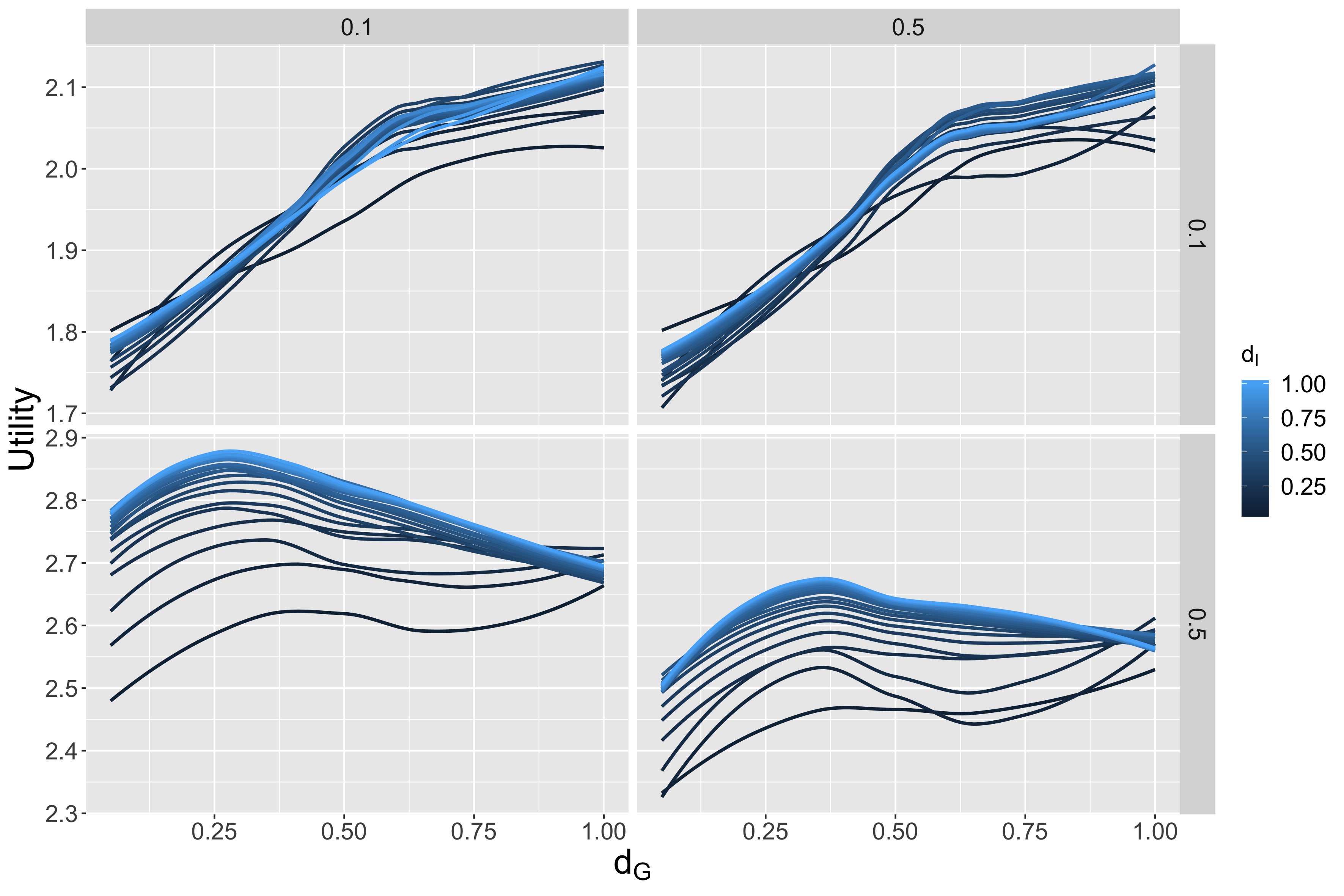}
	\includegraphics[width=0.49\linewidth]{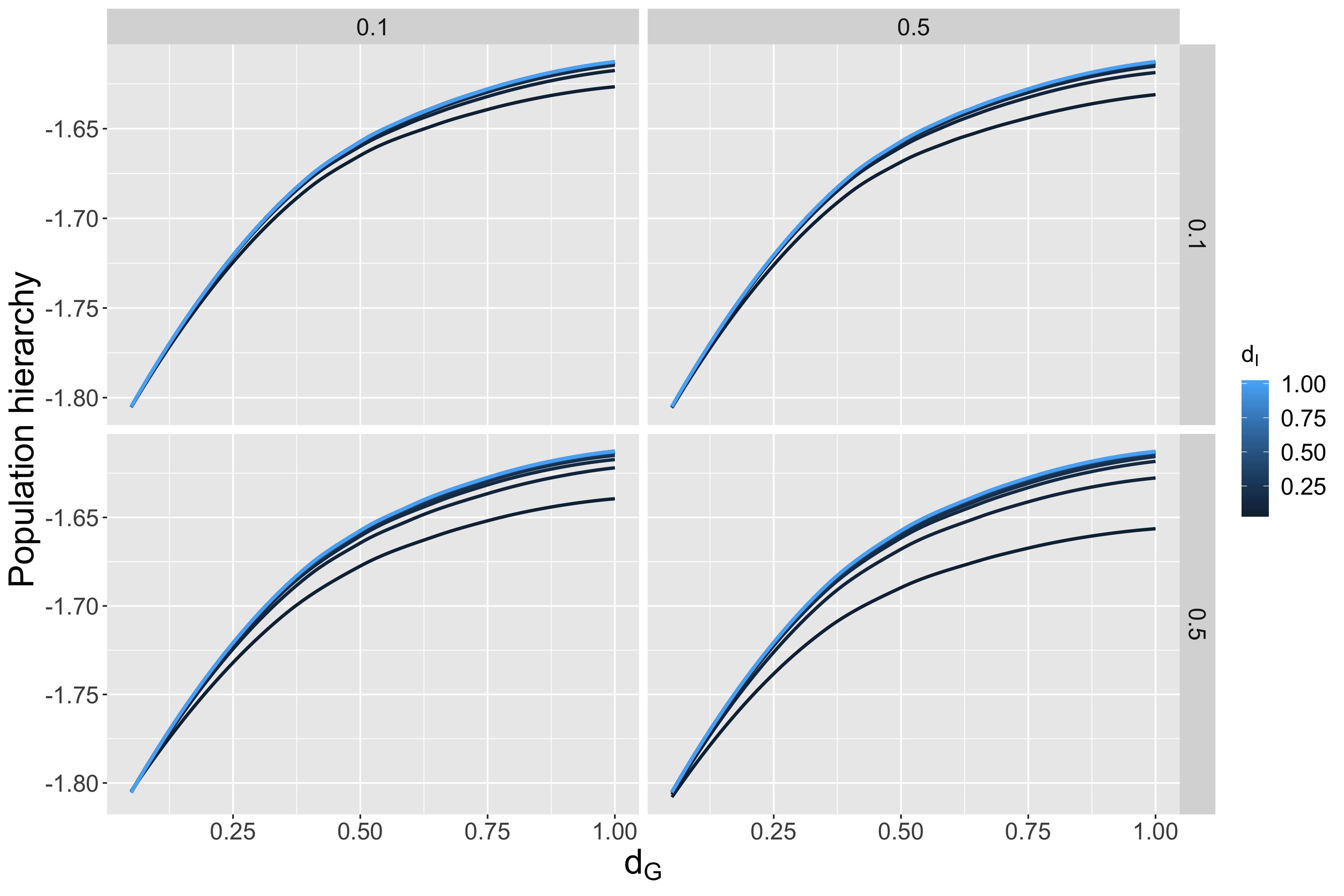}
	\caption{Values of indicators obtained with the grid model exploration. Each plot give each indicator among diversity $D$, utility $U$, innovation $I$ and population hierarchy $\alpha_P$, at fixed parameters $\alpha_I = 1$, $\sigma_U = 1$ and for the log-normal utility law. Indicators are plotted as a function of gravity interaction range $d_G$, for varying values of innovation diffusion range $d_I$ (color). Sub-panels show varying values of mutation rate $\beta$ (rows) and of early adoption rate $r_0$ (columns).\label{fig:exploration}}
\end{figure*}

\paragraph{Model parameters}

Parameter explored in experiments are the spatial interaction parameters $d_G,d_I \in \left[0;2\right]$ which correspond to the crossover mechanism; and the innovation parameter which correspond to the mutation mechanism: innovation rate $\beta \in \left[0;1\right]$, innovation hierarchy $\alpha_I \in \left[0;2\right]$, innovation utility standard deviation $\sigma_U \in \left[ 0.7;2\right]$ (the lower bound is a constraint for the existence of the log-normal with log mean 0), initial penetration rate $r_0 \in \left[0.1;0.9\right]$, and type of innovation utility as normal or log-normal.

\section{Results}

The model is implemented in \texttt{scala} and integrated into the OpenMOLE software \citep{reuillon2013openmole} for numerical experiments. OpenMOLE allows embedding models in any language as black boxes, provides a transparent access to high performance computing infrastructures, and model exploration and validation methods such as sensitivity analysis, design of experiments, and calibration methods. Experiments are designed through workflows using a Domain Specific Language \citep{passerat2017reproducible}.

Source code of the model and analysis, and results are available on the open git repository of the project at \url{https://github.com/JusteRaimbault/UrbanEvolution}. Simulation data files are available on the dataverse repository at \url{https://doi.org/10.7910/DVN/Q5GKZ0}.

\subsection{Internal validation}


As the model is stochastic through the mutation process, first experiments are needed to assess the ``internal validity'' of the model, i.e. to what extent studied indicators are robust to noise and how much stochastic repetitions are needed to significantly distinguish between central values of distributions for different parameter points. We sample thus 100 parameter points using a Latin Hypercube Sampling, and run 1000 model repetitions for each point, in order to estimate the statistical properties across different points in the parameter space.

The sharpe ratios estimated on repetitions as sample standard deviation relative to absolute sample mean have high values for all indicators (min. 3.9 and median 12.1 for diversity; min. 3.0 and median 6.2 for innovation; min. 1.7 and median 3.6 for utility; min. 26 and median 257.3 for population hierarchy), what means that stochastic noise is not an issue for interpreting indicator values.

We also study the distance between points averages relative to their standard deviation, in order to know the significance in comparing averages. This distance is defined by $\Delta_{ij} = 2\frac{\left|\mu_i - \mu_j \right|}{\sigma_i + \sigma_j}$ if $\mu_i$ are estimated central values (mean or median) and $\sigma_i$ estimated standard deviations. These distances have an average and median larger than one for all couples, and a first quartile larger than one for innovation and population hierarchy, while the first quartile is 0.5 for utility and 0.6 for diversity, what means that these two last indicators need more statistical precision to be interpreted. As in the case of normal distributions, a 95\% confidence interval of size $\sigma_2$ is achieved for $n\simeq 64$ runs, a number of 50 runs is satisfying for further experiments.

\subsection{Model exploration}

We then explore a grid of the parameter space consisting of 23,168 parameter points, with a finer step on spatial parameters $d_G$ and $d_I$, and with 50 model replications for each parameter point. We find that the type of distribution for utility has small effects for all indicators and across most parameter values, but particular cases are more interesting in the log-normal case, which we comment further. Similarly, the effect of innovation hierarchy $\alpha_I$ and of innovation utility standard deviation $\sigma_U$ do not induce strong qualitative differences, so we discuss model behavior obtained at $\alpha_I = 1$ and $\sigma_U = 1$.

The Fig.~\ref{fig:exploration} shows the behavior of indicators with remaining parameters varying, namely spatial interaction parameters $d_G$ and $d_I$, mutation rate $\beta$ and innovation adoption $r_0$. Regarding the behavior of diversity (top left panel of Fig.~\ref{fig:exploration}), we find that it always increase for larger population growth spatial interaction distances, meaning that broader population exchanges will induce a higher diversity in innovations, probably through the higher innovation dynamics induced. This increase however reaches a plateau around half of the width of the world ($d_G = 0.5$) which is more obvious for higher values of mutation rates (bottom raw): higher intrinsic innovation dampens the role of spatial interactions. An increase of innovation interaction distance $d_I$ on the contrary monotonously decreases diversity, consistently with the idea of local innovation niches which can be overridden by competing innovations of higher utility coming from further away. Finally, mutation rate has a strong quantitative impact on total diversity as expected, while higher early adoption rate will change diversity behavior only when mutation rates are also high.

\begin{figure}[t]
	\centering
	\includegraphics[width=\linewidth,trim={6cm 6cm 0 6cm}]{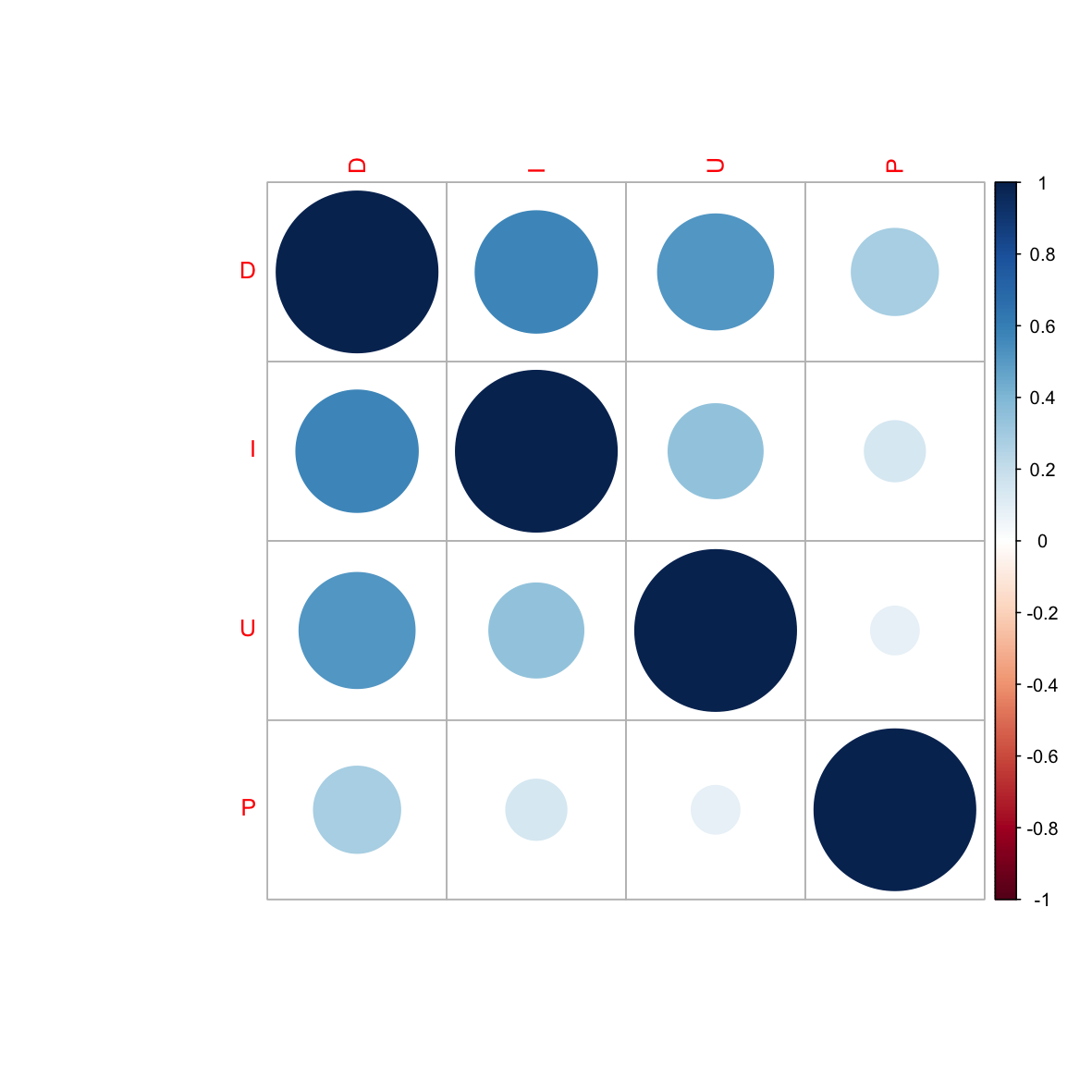}
	\caption{Correlation matrix between indicators, estimated on the full grid experiment. Confidence intervals for correlations estimated with the Fisher method are tiny and not distinguishable in the plot.\label{fig:cormat}}
\end{figure}

Two indicators which behavior is less rich are innovation and population hierarchy (top right and bottom right panels of Fig.~\ref{fig:exploration}), which could be expected as population dynamics are deterministic and strongly related to space, while innovation is directly related to population. We find that these increase both with $d_G$ and $d_I$, meaning that more interactions and more diffusion will foster local innovation. Regarding the final distribution of population, it means that systems where interaction are more global will be less unequal. This is however not always the case in such urban dynamics model as explored by \cite{raimbault2020hierarchy}. The mutation rate $\beta$ here only fosters the role of small $d_I$, yielding higher hierarchies and less innovations for these. The effect of initial adoption rate is small for these indicators. Note that although they appear highly correlated in this region of the parameter space, they are not the indicators with the highest correlation as shown further (see Fig.~\ref{fig:cormat}).

Finally, the indicator with a more interesting behavior is the utility (bottom left panel of Fig.~\ref{fig:exploration}). For low values of mutation rates, with find a piecewise linear behavior as a function of $d_G$ and small effects of $d_I$, meaning that low innovation settings yield regimes where broader interactions are systematically desirable for the all system. However, when $\beta$ is higher, we witness a maximum of utility as a function of $d_G$, consistent across different values of $d_I$ and of $r_0$. This corresponds to an intermediate regional regime where local innovation regimes are more beneficial than a global integration. This peak is the strongest when innovation diffuse at broader range, what may mean that this regime is due to the emergence of regional ensemble of comparable competitivity and thus a higher chance of high utility innovation, while a globalized system would concentrate on a single dominating city and be overall less performant.

We also compute the correlation matrix between indicators across the full grid experiment. This shows that correlations that one could visually extrapolate from studying a part of the parameter space as commented in Fig.~\ref{fig:exploration} do not correspond to the actual correlations on the broader parameter space. Note that local correlation matrices could be estimated for a more thorough discussion. We show the correlation matrix in Fig.~\ref{fig:cormat}. Confidence intervals estimated with the Fisher method are of negligible width compared to correlation values. We find that diversity is finally the indicator with highest correlations. The correlation between innovation and population is low, meaning that the similar curves observed before are a particular case. Innovation and utility have a correlation lower than 0.5, and are thus rather independent. Regarding the effective dimension of the indicator space, a principal component analysis on normalized indicators gives 48\% of variance on the first component, 77\% of cumulated variance on the second and 91\% on the third, confirming that even if correlation exist the behavior of indicators are rather independent.

\subsection{Model optimization}

\begin{figure}[t]
	\centering
	\includegraphics[width=\linewidth]{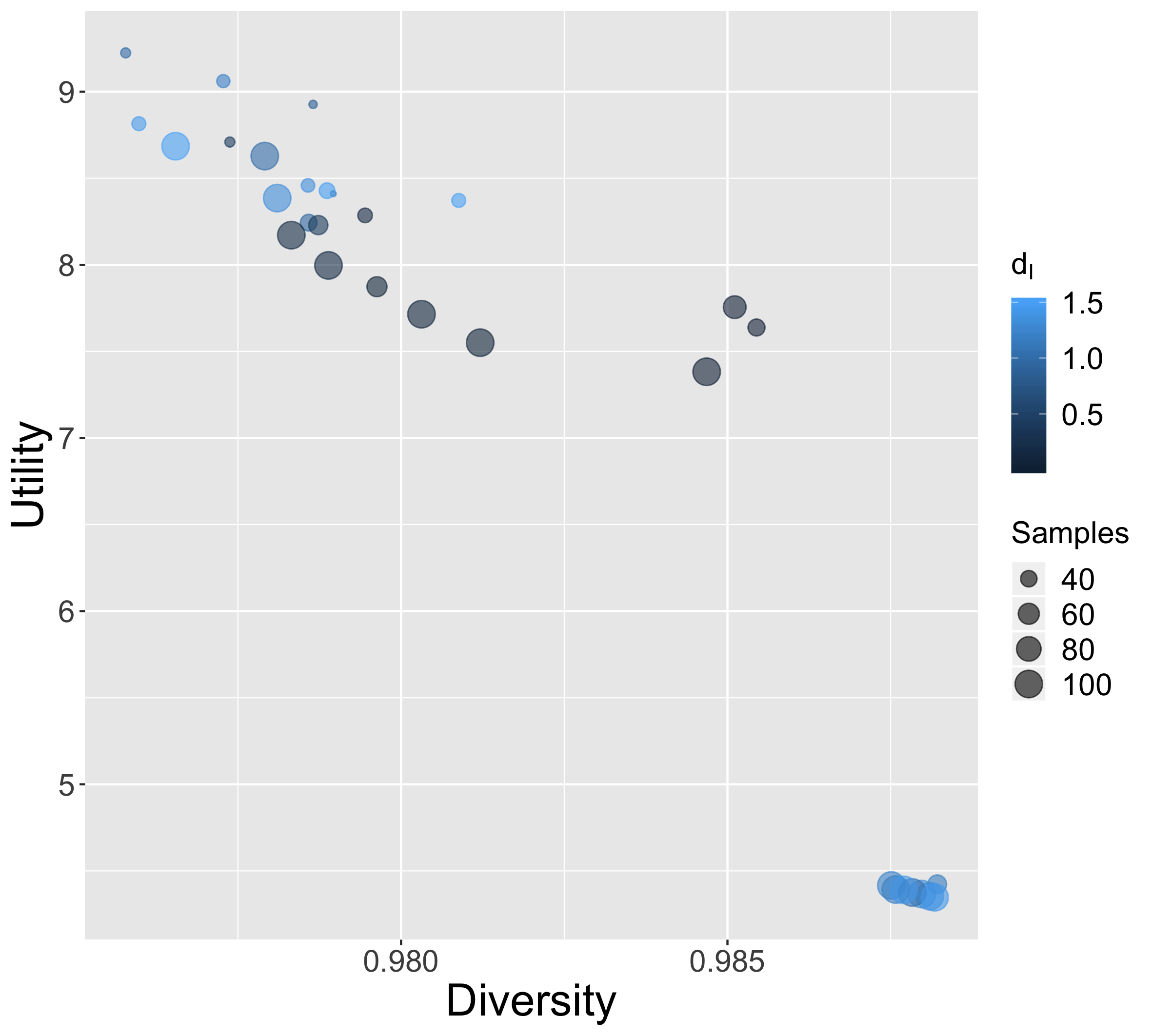}
	\caption{Pareto front for the contradictory indicator of utility and diversity obtained with the NSGA2 algorithm. Point color gives innovation interaction range $d_I$ while point size gives the number of stochastic samples.\label{fig:pareto}}
\end{figure}

The last experiment we perform is the application of a multi-objective optimization procedure to the model. More particularly, one could want to optimize simultaneously the global utility but also the diversity to ensure a certain resilience in the urban system. We apply thus a bi-objective genetic algorithm on the model, trying to maximize simultaneously utility and diversity. The algorithm is the NSGA2 algorithm \citep{deb2002fast}, which we run for 10,000 generations with a population of 200 individuals on a Island distribution scheme. The algorithm is integrated into the OpenMOLE software, and the stochastic aspect is internally tackled by using an embedding strategy, i.e. by adding the number of repetitions as an additional objective to find compromise points between the number of run needed and their statistical accuracy.

We show in Fig.~\ref{fig:pareto} the obtained points, which are close to a Pareto front. We find two regimes, the first corresponding to the upper points in the plot for higher utilities but lower diversities, which should correspond to the intermediate optimal regimes identified before, and in which some kind of linear compromize between utility and diversity exists: increasing global utility is done at the cost of reducing diversity. Within this first regime, two very different parameter setting coexist, one with high innovation diffusion (light blue, points with higher utility), the other in dark blue with a more local innovation diffusion. This means that reducing the span of innovation diffusion will increase diversity as one could expect. A second part of the Pareto front, obtained after a sharp transition, allows increasing the diversity but at the price of a much lower utility (points on the bottom right). These points are all close to equivalent, and one may prefer in the compromise the points just before the transition which correspond to a local diffusion setting. Thus, local diffusion correspond to intermediate compromises, while broad diffusion corresponds to extremes in the Pareto front. In a nutshell, this optimization exercise is interesting both to show how the model produces compromises, but also how it could be used in practice for innovation policies applied to systems of cities.

\section{Discussion}

Our simple formulation of an urban genome and associated evolutionary processes, implemented by the diffusion of innovation, already capture complex urban dynamics, as shown for example by the emergence of an intermediate optimal regime corresponding to local innovation niches which diffuse far. Although much more empirical work would be needed to compare these stylized facts to real world settings, our model suggests in this particular setting that too much integration is not always optimal, what corresponds to the theoretical fact that complex systems are generally modular at different scales \citep{ethiraj2004modularity}. More generally, regarding the implications of our results for possible formalizations and theories of urban evolution which would closely build on biological and cultural evolution by extending them, we have demonstrated how a particular instance of an urban genome can be used to simulated urban dynamics, including fundamental processes needed to effectively have evolution. To what extent this approach relates to existing approaches of urban evolution which use other definitions remains to be investigated.

Several extensions and applications would be possible to this first model exploration. First, the innovation space in our model remained unidimensional. \citep{hidalgo2007product} show in terms of industrial production by countries that these industrial spaces are highly dimensional in terms of product produced and interdependencies between countries and types of production, increasing the path-dependency in economic trajectories. The investigation of patent data from a semantic viewpoint also shows the highly dimensional nature of technological innovation \citep{bergeaud2017classifying}. A direct extension of our model would consist in having a matrix genome instead of a vector one. Innovations would occur across several dimensions which can correspond to industrial or technological domains, but also social and cultural innovation and innovations in terms of infrastructures which condition the way systems of cities evolve. Each dimension should have particular innovation rules depending on its nature, and dependencies across dimensions could be introduced, implementing the possible emergence of technological co-evolution niches beside the spatial co-evolution niches.

Second, applying the model to real system of cities, both in terms of initial parametrization and of empirical laws for innovation processes, would allow bringing this approach closer to possible policy applications. \cite{raimbault:halshs-01880492} benchmarked several models of urban growth based on interactions between cities, and integrated the model of \cite{favaro2011gibrat} on which this work was based. It however only included accurate initial populations and distance matrix. This application to real systems of cities also allows testing the performance of the model in predicting possible trajectories for these systems.

Finally, our approach was rather restricted in the sense that even a broad geographical range is taken into account, a single ontological scale is included in the model, i.e. the macroscopic scale since cities are the basic agents. Similarly, a single temporal scale was included, although two dynamics of spatial interaction and innovation diffusion are effectively combined. A multi-scale approach of urban evolution would be necessary to fully account for the complexity of these systems. \cite{raimbault:halshs-02351722} introduced a multi-scalar model for population growth with upward and downward strong feedbacks between the mesoscopic scale (urban form) and the macroscopic scale (spatial interaction model). Adapting this approach in the case of urban evolution through innovation diffusion would be an interesting potential development.

\section{Conclusion}

We have introduced a simple model of urban evolution integrating effectively the evolutionary processes of transmission, transformation and evoluting sub-systems, through innovation diffusion and spatial interactions. Model exploration yield complex behavior while multi-objective optimization shows the potentiality for the model to produce compromises between utility and diversity in the system of cities. This work is thus a first step towards more elaborated models of urban evolution.

\bigskip

\section{Acknowledgements}

Results obtained in this paper were computed on the vo.complex-system.eu virtual organization of the European Grid Infrastructure ( http://www.egi.eu ). We thank the European Grid Infrastructure and its supporting National Grid Initiatives (France-Grilles in particular) for providing the technical support and infrastructure. This work was funded by the Urban Dynamics Lab grant EPSRC EP/M023583/1.

\bigskip

\footnotesize

\end{document}